\documentclass[prb,aps,nobibnotes,showpacs,twocolumn]{revtex4}
\usepackage{graphicx}%
\usepackage{dcolumn}
\usepackage{amsmath}   
\usepackage{bm}
\begin{document}

\title{\centering\Large\bf Paraelectric and ferroelectric order in 
                           two-state dipolar fluids}
\author{Dmitry V.\ Matyushov}
\email[E-mail:]{dmitrym@asu.edu.}
\author{Andriy Okhrimovskyy}

\affiliation{ 
Department of Chemistry and Biochemistry, Arizona State University, PO Box 871604, Tempe, 
AZ 85287-1604}

\date{\today}

\begin{abstract}
  Monte Carlo simulations are used to examine cooperative creation of
  polar state in fluids of two-state particles with nonzero dipole in
  the excited state. With lowering temperature such systems undergo a
  second order transition from nonpolar to polar, paraelectric phase.
  The transition is accompanied by a dielectric anomaly of
  polarization susceptibility increasing by three orders of magnitude.
  The paraelectric phase is then followed by formation of a nematic
  ferroelectric which further freezes into an \textit{fcc}
  ferroelectric crystal by a first order transition. A mean-field
  model of phase transitions is discussed.
\end{abstract}
\preprint{Typesetting using REV\TeX 4} 

\maketitle 

In this Letter we describe Monte Carlo (MC) simulations of a fluid
composed of two-state (TS) soft sphere (SS) particles interacting with
the dipole-dipole potential. This model system, referred to as TS/SS
fluid, shows a complex phase diagram including the transition from a
nonpolar to polar, paraelectric phase followed by the transition to
the ferroelectric phase. The existence of ferroelectric order in
dipolar systems was first suggested by Debye\cite{Debye:12} who
predicted a transition at $y_c=1$; $y=(4\pi/9) m^2\rho/kT$, where $m$ is
the dipole moment and $\rho$ is the number density.
Onsager\cite{Onsager:36} and Kirkwood\cite{Kirkwood:39} have shifted
the transition temperature to zero, $y_c\to \infty$. The generalized
mean-field theory by H\o ye and Stell\cite{Hoye:76} predicts
$y_c=(1-\Theta)^{-1}$, $0\leq \Theta\leq 1$ thus allowing transition at any
temperature between the Debye and Onsager-Kirkwood limits.  Recent
computer simulations have indicated that isotropic ferroelectric phase
is possible for polar fluids at a non-zero
temperature,\cite{Wei:92,Ayton:96,Gao:00} although the issue of
boundary conditions is still debated.\cite{Wei:93} Most of the
discussion of spontaneous order in dipolar systems has focused on
systems with permanent dipoles. Real systems, either of molecular or
nano-scale dimension, are composed of polarizable particles. The
present simulations establish the existence of spontaneous
ferroelectric order in polarizable dipolar fluids.

Many strongly dipolar states are created by intramolecular separation
of charge, either thermal or optical, between the the donor and
acceptor parts of a molecule. Such states, common in chemistry and
biology,\cite{Marcus:93} often behave as independent donor-acceptor
complexes with no cooperativity of charge-transfer transitions.  A
dramatic distinction from this situation is the Peierls instability in
organic charge-transfer salts where charge-separated states are
cooperatively created in one-dimensional stacks of donor and acceptor
units.\cite{Soos:74} It appears that there is no fundamental reason
why this type of transition should be limited to the crystal phase.
We show that the cooperative coupling between TS dipoles leads to the
nonpolar/polar second-order phase transition in the liquid phase.

The Hamiltonian of a fluid of TS/SS particles is
\begin{equation}
  \label{eq:1}
  H= \sum_j H_{TS}(j) + \sum_{j<k} \left(u_{SS}(jk) - \hat n_j \mathbf{m}_j\cdot \mathbf{T}_{jk} \cdot \mathbf{m}_k \hat n_k\right).
\end{equation}
Each individual particle $j$ is characterized by the vacuum two-state
Hamiltonian $H_{TS}(j)$ with the excited state population $\hat n_j$, the
vacuum energy gap $\Delta I$ and the mixing between the states $V$.  The
dipole moment is zero in the vacuum ground state and is $\mathbf{m}$
in the excited state. The excited-state dipoles interact via the
dipole-dipole potential with
$\mathbf{T}_{jk}=-\nabla_j\nabla_k|\mathbf{r}_j-\mathbf{r}_k|^{-1}$ in Eq.\ 
(\ref{eq:1}). Finally, the SS repulsion is
\begin{equation}
  \label{eq:3}
  u_{SS}(jk)=4\epsilon\left(\sigma/r_{jk}\right)^{12} .
\end{equation}
The model is characterized by the reduced density, $\rho^*=\rho\sigma^3$,
reduced temperature, $T^*=kT/ \epsilon$, and reduced dipole moment,
$m^*=m/\sqrt{\epsilon\sigma^3}$. The reduced parameters for the TS system are:
$V^*=V/kT$ and $I^* = \Delta I/ kT$.  The intermolecular interactions are
fully characterized by two parameters: $x=(\rho^*)^4/T^*$
for SS repulsions\cite{Hansen:70} and $y$ for dipolar coupling.

The exact calculation of the energies of $N$ particles requires
diagonalization of the $(2N)^2$ matrix. This is still prohibitively
slow for condensed phase simulations.  We will therefore use the
Hartree decoupling assuming that the field acting on a given particle
is produced by average excited-state populations of other
particles.\cite{Winn:92} The ground state energy of the fluid $E=\langle \prod_j
\Psi_{gj}|H|\prod_k \Psi_{gk}\rangle$ is then defined on the wave functions $\Psi_{gj}$
diagonalizing the Hamiltonian
\begin{equation}
  \label{eq:2}
  H(j) = H_{TS}(j) - \hat n_j \mathbf{m}_j\cdot \mathbf{R}_j ,
\end{equation}
where 
\begin{equation}
\label{eq:71}
\mathbf{R}_j = \sum_k \mathbf{T}_{jk}\cdot\mathbf{m}_k \langle n_k\rangle 
\end{equation}
is the reaction field of the system dipoles. One gets
\begin{equation}
  \label{eq:4}
  E = \sum_j E_{g}(j) + \sum_{j<k} (u_{SS}(jk) - u_{DD}(jk)).
\end{equation}
Here,  the energy of each particle is
\begin{equation}
  \label{eq:5}
  E_{g}(j) = \frac{1}{2}\left[\Delta I + u_{Rj} - \Delta E_j\right]
\end{equation}
and the energy of the ground state in vacuum is set to be zero of
energy.  The energy gap $\Delta E_j = \left[(\Delta I + u_{Rj})^2 + 4V^2
\right]^{1/2}$ in Eq.\ (\ref{eq:5}) is modulated by the coupling of
the excited-state dipole $\mathbf{m}_j$ to the reaction field, $u_{Rj}
= - \mathbf{m}_j\cdot\mathbf{R}_j$.  Finally, the excited-state population
$2\langle n_j\rangle = 1 - (\Delta I + u_{Rj})/\Delta E_j$ defines the dipole moment
$\bm{\mu}_j=\mathbf{m}_j\langle n_j\rangle $ in the dipole-dipole interaction
potential in Eq.\ (\ref{eq:4}): $u_{DD}(jk)= - \bm{\mu}_j \cdot
\mathbf{T}_{jk}\cdot \bm{\mu}_k $.  The dipole moment of each particle thus
fluctuates with the instantaneous value of the reaction field.

The MC simulations of TS/SS fluids were performed for the $NVT$
ensemble of $N=256$ particles in a cubic box with periodic boundary
conditions. The long range dipolar interactions were accounted for by
the reaction field method with infinite reaction field dielectric
constant.  An MC trial move combines displacement and rotation of a
particle followed by iterative, self-consistent calculation of the
fields $\mathbf{R}_j$.\cite{Vesely:76} Each trial move thus results in
a new set of dipoles $\bm{\mu}_j$. This calculation is the most
time-consuming part of the simulation protocol.  Typical simulations
were 8$\times10^5$ cycles long (170 h on a single Alpha/740MHz processor),
points close to phase transitions required simulations with 1.2$\times10^6$
cycles.

The appearance of the nonpolar/polar (NP) and polar/ferroelectric (PF)
transitions was monitored by calculating the polar order parameter $ P
= \langle\mu\rangle /m = (Nm)^{-1} \langle \sum_j \mu_j \rangle $ and the usual orientational
first-order (ferroelectric) and second-order (nematic) parameters,
$S_1$ and $S_2$, respectively.\cite{Wei:92,Allen:96} The isotropic,
nonpolar phase is characterized by all order parameters equal to zero.
The polar state is marked by $P\neq0$, whereas $P$, $S_1$, and $S_2$ are
all non-zero in the ferroelectric nematic.

\begin{figure}[htbp]
  \centering \includegraphics*[width=5.5cm]{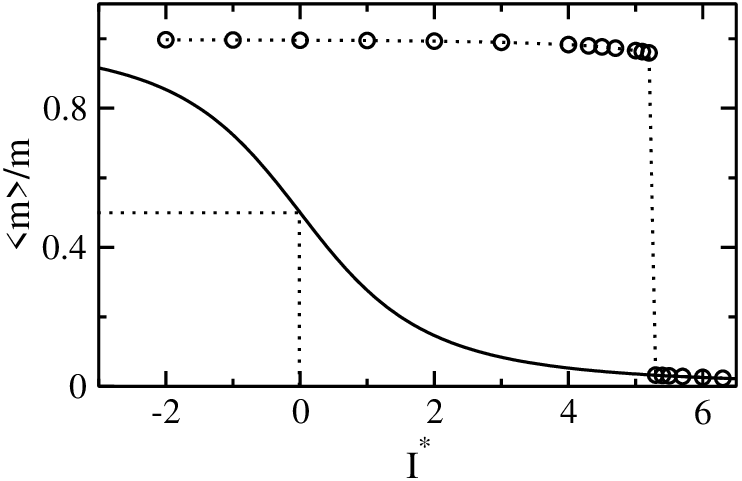}
  \caption{Polar order parameter of non-interacting particles (line) 
    and of the TS/SS fluid (points) vs the reduced vacuum energy gap
    $I^*=\Delta I/kT$; $\rho^*=0.8$, $m^*=2.0$, $T^*=0.8$. The dashed line
    indicate the point $\langle\mu \rangle/m=1/2$ at $I^*=0$,
    the dotted line connects the simulation points.}
  \label{fig:1}
\end{figure}

Spontaneous creation of a non-zero average dipole moment is a result
of cooperative coupling between induced dipoles.  The dipole moment
$\mu$ of an isolated particle changes smoothly with the energy gap $I^*$
reaching one half of its maximum value $m$ at zero gap $I^*=0$ (Fig.\ 
\ref{fig:1}). This point corresponds to a purely covalent state of the
donor-acceptor complex. The ionic character, $\mu/m \simeq 1$, is
achieved at $I^*<0$, $|I^*/V^*| \gg 1$. In contrast, when the
particles are allowed to interact in the fluid phase, cooperativity
results in a very sharp change of $\langle\mu\rangle$ from zero to $m$ at the
transition value $I_{NP}^* >0$ (Fig.\ \ref{fig:1}, $I_{NP}^* = 5.2$ at
$\rho^*=0.8$, $T^*=0.8$, $m^*=2.0$, and $V^*=-1.0$). Thermal fluctuations
of induced dipoles thus reinforce each other leading to a negative
energy of interaction with the reaction field, $\langle u_R\rangle <0$, which
overrides the positive vacuum energy gap and the polarization energy
invested in creating the dipole. 

This type of transition was predicted by Logan\cite{Logan:86,Logan:87}
who suggested that a solution of alkali atoms can show a continuous
transition from a normal insulator (non-polar) to excitonic insulator
(paraelectric) phase.\cite{Hall:86,Xu:88,Winn:92} Although the atomic
dipole is created by hybridizing the non-polar atomic states through the
transition dipole, the basic physics of reaction-field stabilization
overriding the energy gap\cite{Fois:94,Spezia:03} is common to the
present problem and the excitonic insulator transition. In is
interesting to note that non-linear polarizability effects
characteristic of the two-state description might be important for the
transition to occur since the complex phase diagram obtained here for
the TS/SS fluid does not exist for the fluid of Drude oscillators
often used to model atomic and molecular polarizability.\cite{DMjpca:04}

\begin{figure}[htbp]
  \centering
  \includegraphics*[width=5.5cm]{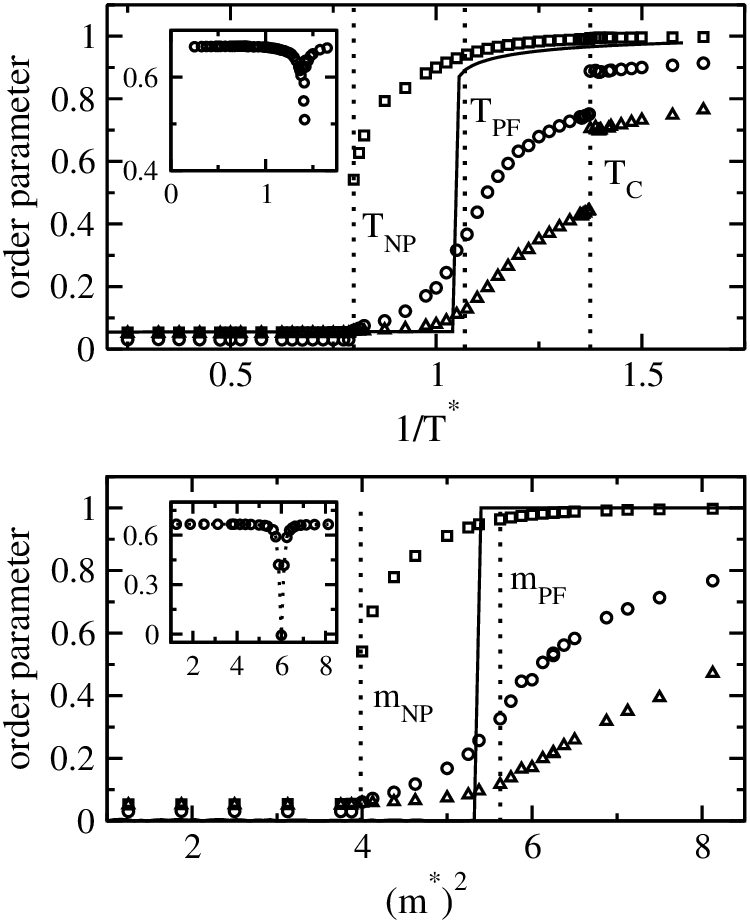}
  \caption{Order parameters $P$ (squares), $S_1$ (circles), and $S_2$ (triangles) 
    of TS/SS fluids vs $1/T^*$ (upper panel, $m^*=2.0$) and vs
    $(m^*)^2$ (lower panel, $T^*=1.25$) at $\rho^*=0.8$, $I^*=4.0$,
    $V^*=-1.0$. The solid lines show the results of mean-field theory
    calculations [Eqs.\ (\ref{eq:11})--(\ref{eq:14})] for the order
    parameter $P$. The Binder parameter,\cite{Binder:92} $1 - \langle E^4\rangle
    /3\langle E^2\rangle^2$, vs $1/T^*$ (upper panel) and vs $(m^*)^2$ (lower
    panel) is shown in the insets. }
  \label{fig:2}
\end{figure}

\begin{figure}[htbp]
  \centering \includegraphics*[width=5.5cm]{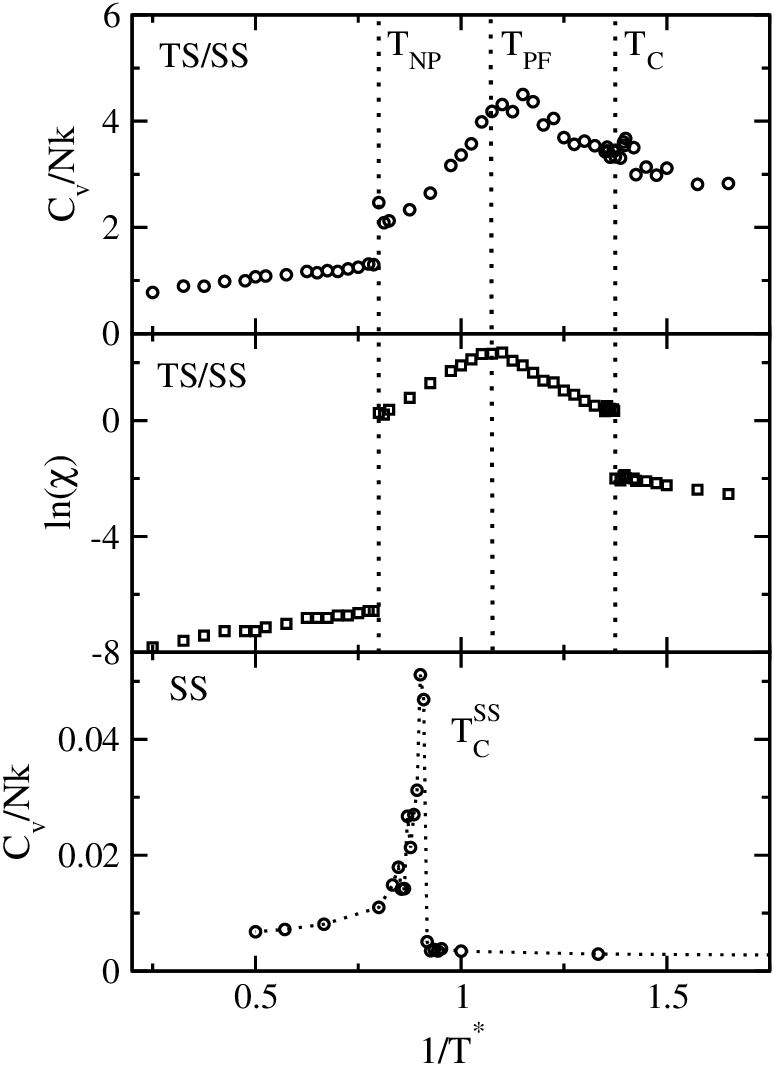}
  \caption{Heat capacity $C_v$ (upper panel) and polarization 
    susceptibility $\chi$ 
    (middle panel) of the TS/SS fluid and heat capacity of the SS
    fluid (bottom panel).  $m^*=2.0$, $I^*=4.0$, $V^*=-1.0$ for the
    TS/SS fluid, $\rho^*=0.8$ for both the TS/SS and SS fluids.  MC
    simulations of the SS fluid were performed with $N=2048$ particles
    in the box. The typical simulation length was 50 000 cycles. }
  \label{fig:3}
\end{figure}

The MC simulations shown below consider the variation of
intermolecular repulsion and attraction through $T^*$ and $m^*$ at
constant $\rho^*$, $I^*$, and $V^*$.  Three order parameters as functions
of $T^*$ and $m^*$ are shown in Fig.\ \ref{fig:2}.  The parameter $P$
increases sharply at $T_{NP}^*/ m_{NP}^*$ marking the onset of the
polar (paraelectric) phase.  Both $S_1$ and $S_2$ start to increase
with decreasing temperature or increasing dipole moment resulting in
the PF transition at $T_{PF}^*/m_{PF}^*$. The latter point is defined
from the peak of dielectric susceptibility (Fig.\ \ref{fig:3}, the
simulation data corresponding to the change of $m^*$ at constant $T^*$
are not shown). A similar susceptibility peak was observed for the PF
transition in a fluid of hard sphere Ising spins with a square-well
attraction.\cite{Ballone:87}

Finally, both $S_1$ and $S_2$ have a discontinuous jump at
$T_C^*=0.73$.  The analysis of the density structure factors and pair
distribution functions indicates that the system is in fluid phase
above $T_C$ and freezes into the \textit{fcc} ferroelectric at $T_C$.
Further, the phase between $T_{PF}$ and $T_C$ is a ferroelectric
nematic as is seen from the analysis of angular projections of the
pair distribution function and from the analysis of the position
distribution functions parallel and perpendicular to the director (not
shown here).  These functions show no indication of spatial structure
thus ruling out a possibility of smectic order. No crystallization was
achieved by changing $m^*$ at constant $T^*$ in the range of values
studied.

\begin{figure}[htbp]
  \centering \includegraphics*[width=5.5cm]{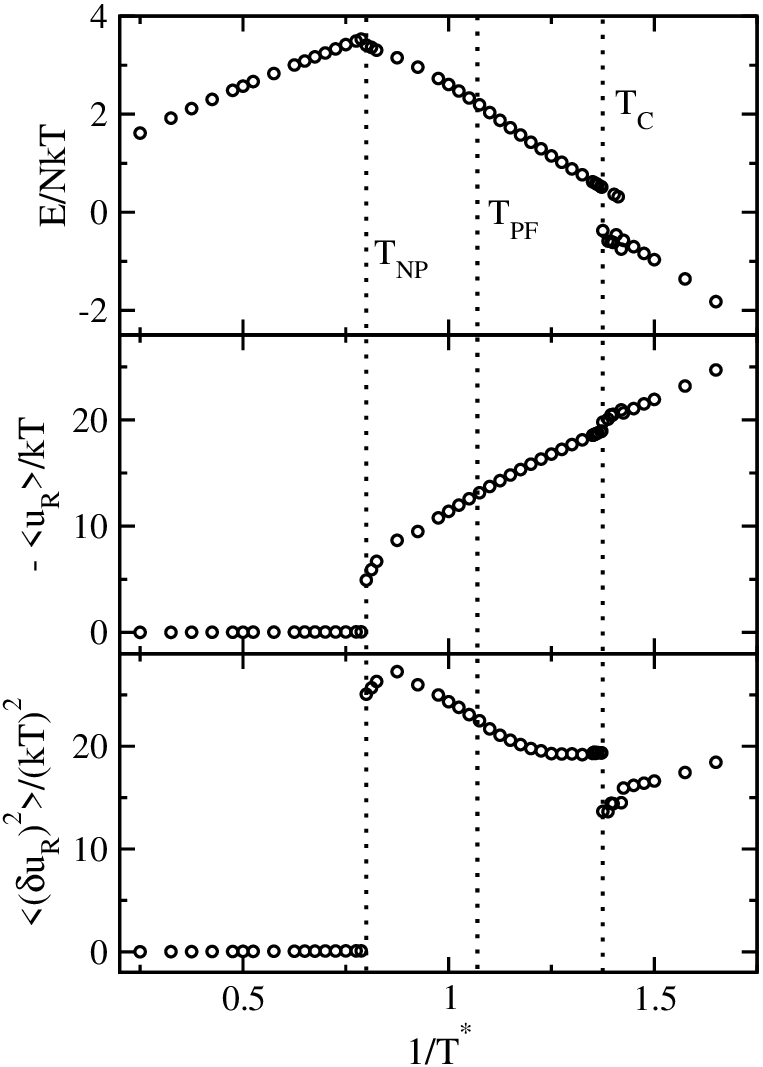}
  \caption{Total average energy per particle, $E/NkT$, the energy of dipolar coupling to 
    the reaction field, $-\langle u_R\rangle$, and its variance,
    $\langle(\delta u_R)^2\rangle$, vs $1/T^*$ for the TS/SS fluid:
    $\rho^*=0.8$, $m^*=2.0$, $I^*=4.0$, $V^*=-1.0$.}
  \label{fig:4}
\end{figure}

The NP transition is characterized by discontinuity of the heat
capacity $C_v/Nk=\langle(\delta E)^2\rangle/N(kT)^2$ and a very substantial (3 orders
of magnitude) increase in the dielectric susceptibility $\chi=\rho \langle(\delta
\mathbf{M})^2\rangle /NkT$, where $\delta\mathbf{M}$ is the fluctuation of the
total dipole of the system of $N$ particles (Fig.\ \ref{fig:3}). At
the same time, the total energy is continuous pointing to a
second-order phase transition (Fig.\ \ref{fig:4}).  The PF transition
is continuous both in energy and heat capacity and is manifested by
maxima of $C_v$ and $\chi$ (Fig.\ \ref{fig:3}).  Finally, crystallization
is a first-order transition with the latent heat arising mostly from
the change in the dipolar interaction energy.  The continuous
character of the NP and PF transitions is supported by the value of
the Binder parameter,\cite{Binder:92} $1 - \langle E^4\rangle /3\langle E^2\rangle^2$,
which is very close to 2/3 (continuous transition) in the entire range
of parameters except at crystallization when it takes a dip consistent
with the first order of this transition.  The onset of the polar phase
is also marked by a continuous increase in the average reaction filed
and a discontinuous jump in its variance (Fig.\ \ref{fig:4}, bottom
panel).  The breakdown of the linear response approximation, $-kT\langle
u_R\rangle=\langle(\delta u_R)^2\rangle$, at the onset of paraelectric phase\cite{Spezia:03}
is characteristic of polarizable systems.\cite{DMjpca:04}

The temperature of NP transition falls in the region of freezing
transition of the reference fluid with the repulsive potential
$U_{SS}=\sum_{j<k}u_{SS}(r_{jk})$ which is characterized by a peak of the
heat capacity $C_v/Nk = \langle (\delta U_{SS})^2\rangle/N(kT)^2 $ (bottom panel in
Fig.\ \ref{fig:3}). The SS fluid crystallizes into an \textit{fcc}
lattice below the transition temperature $T^{SS}_C$ as we found from
the density structure factors in accord with previous reports in the
literature.\cite{Laird:90} It may be suggested that positional
instabilities of the reference repulsive potential drive the PF
transition of the TS/SS fluid.  However, dipolar interactions do not
favor the highly symmetric \textit{fcc} lattice,\cite{Groh:01} and the
system stays in the fluid phase in the temperature range $T_{PF}\leq T\leq
T_C$ crystallizing at $T_C$. This interpretation is consistent with
the notion advocated by several authors\cite{Groh:01} that dipoles
suppress freezing into a high-symmetry lattice due to the anisotropic
nature of dipolar forces. Note that the \textit{fcc} lattice does not
necessarily represent the ground state since a ferroelectric solid can
never be strictly cubic. Other stable structures may be suppressed in
simulations by the cubic shape of the simulation cell and periodic
boundary conditions.\cite{Gao:00}

The energy per particle $e=E/NkT$ depends on the fluctuating variable
$u=u_R/kT$. The mean-field solution, neglecting these fluctuations,
is given as
\begin{equation}
  \label{eq:11}
  \langle e \rangle = (P/2)\langle u\rangle - P \Delta e(\langle u\rangle),
\end{equation}
where $\Delta e =\Delta E/kT$. The average $\langle u\rangle$ is affected by the macroscopic
reaction field 
\begin{equation}
 \label{eq:R0}
\mathbf{R}_0=(4\pi/3)m\rho PS_1(1-\Theta)\mathbf{\hat d} 
\end{equation}
caused by spontaneous ferroelectric polarization of the liquid and the
microscopic reaction field $\mathbf{R}_{\text{mic}}$ due to
dipole-dipole correlations. Here, $\Theta$ is the mean-field parameter
of H\o ye and Stell\cite{Hoye:76,Hoye:95} correcting the macroscopic
field by the account of local dipolar correlations. If one assumes
that the microscopic correlations are not affected by the macroscopic
order, $\langle u\rangle$ becomes
\begin{equation}
  \label{eq:12}
   \langle u\rangle = 3yP(1-\Theta)S_1^2 - (2/P) u_D(\langle\mu\rangle),
\end{equation}
where $u_D(\langle\mu\rangle)=-\beta\langle\mathbf{\mu}\rangle\cdot\mathbf{R}_{\text{mic}}/2$ is the
reduced average interaction energy between particles carrying the
average dipole $\langle\mu\rangle=Pm$. Such dipolar interactions are well described
by the Pad{\'e}-truncated perturbation expansion\cite{Gubbins:84}
resulting in the following free energy per particle
\begin{equation}
  \label{eq:13}
   f_{D}/kT= -\frac{y^2P^4 I_2(x)}{1 + yP^2 I_3(x)/I_2(x)} .
\end{equation}

The perturbation integrals in Eq.\ (\ref{eq:13}) are defined on the
reference SS fluid as
\begin{equation}
  \label{eq:14}
  \begin{split}
    I_2(x) & = (27/128\pi^2)\langle(\rho^2r^6)^{-1}\rangle_{SS} , \\ 
    I_3(x) & = (27/512\pi^3)\langle\rho^3W_{DDD}\rangle_{SS},
  \end{split}
\end{equation}
where $W_{DDD}$ is the Axilrod-Teller potential.\cite{Allen:96} The
average over the configuration of the SS fluid, $\langle\dots\rangle_{SS}$, has
been calculated from MC simulations.  $I_2(x)$ and $I_3(x)$ obtained
from simulations at different temperatures and densities all fall on
one universal dependence on $x$. However, the scaling $I_2(x)\propto
x^{-0.5}$ and $I_3(x)\propto x^{-0.75}$ predicted from the statistical
average is not seen in the simulations. Instead, much weaker power
laws, $I_2(x)=0.1433x^{-0.138}$ and $I_3(x)=0.0314x^{-0.09915}$, fit
the data.  The perturbation integrals obtained at different $\rho^*$ and
$T^*$ are smooth functions of $x$ up to $\rho^*=0.9$.

The paraelectric order parameter can be calculated with the mean-field
solution for $\langle u\rangle$ and the free energy obtained by thermodynamic
integration of $\langle e\rangle $ in Eq.\ (\ref{eq:11}).  The mean-field solution
does not reproduce the separation of the NP and PF transitions and
instead gives a direct transition from a non-polar phase to a polar,
ferroelectric phase at a single transition point. The
calculation shown in Fig.\ \ref{fig:2} is performed for the $\Theta$
parameter in Eq.\ (\ref{eq:R0}) from the mean-spherical solution for
hard dipolar spheres.\cite{Hoye:76,Wertheim:71} The agreement with
simulations for the PF transition is very good for both the dependence
on temperature (upper panel in Fig.\ \ref{fig:2}) and the dependence
on the dipole moment (lower panel in Fig.\ \ref{fig:2}). A model
including fluctuations of the reaction field\cite{Zhang:95} may result
in a more accurate description of the NP and PF transitions yielding a
separate paraelectric phase.

\begin{acknowledgments}
This work was supported by the NSF (CHE-0304694).
\end{acknowledgments}  

\bibliographystyle{apsrev}
\bibliography{/home/dmitry/p/bib/chem_abbr.bib,/home/dmitry/p/bib/et,/home/dmitry/p/bib/liquids,/home/dmitry/p/bib/ferro,/home/dmitry/p/bib/photosynth,/home/dmitry/p/bib/solvation,/home/dmitry/p/bib/dm}

\end{document}